\documentstyle[11pt,newpasp,twoside,epsf]{article}
\begin{document}
\title{High-resolution simulations and visualization of protoplanetary
disks}
\author{Pawe{\l} Cieciel\c{a}g, Tomasz Plewa, Micha{\l} R\'o\.zyczka}
\affil{Nicolaus Copernicus Astronomical Center, Bartycka 18, 00-716 Warsaw,
Poland}
\begin{abstract}

A problem of mass flow in the immediate vicinity of a planet embedded
in a protoplanetary disk is studied numerically in two dimensions.
Large differences in temporal and spatial scales involved suggest that
a specialized discretization method for solution of
hydrodynamical equations may offer great savings in computational
resources, and can make extensive parameter studies
feasible. Preliminary results obtained with help of Adaptive Mesh
Refinement technique and high-order explicit Eulerian solver are
presented. This combination of numerical techniques appears to be an
excellent tool which allows for direct simulations of mass flow in
vicinity of the accretor at moderate computational cost. In
particular, it is possible to resolve the surface of the planet
and to model the process of planet growth with minimal set of
assumptions. Some issues related to visualization of the results and
future prospects are discussed briefly.

\end{abstract}
\section{The Method}
Extremely small temporal and spatial scales involved in the problem of
accretion onto a protoplanet necessitate the use of nonuniform
discretization in the vicinity of the accretor. In our study we used
adaptive mesh refinement (AMR) method combined with a high-resolution
Godunov-type advection scheme ({\sc amra}, Plewa \& M\"uller
2000). The AMR discretization scheme follows the approach of Berger
and Colella (1989). The computational domain is covered by a set of
completely nested {\em patches} occupying {\em levels}. The levels
create a refinement hierarchy. As one moves toward higher levels, the
numerical resolution increases by a prescribed integer factor
(separate for every direction). The net flow of material between
patches at different levels is carefully accounted for in order to
preserve conservation properties of hydrodynamical equations.
Boundary data for child patches are either obtained by parabolic
two-dimensional conservative interpolation of parental data or set
according to prescribed boundary conditions.

Hydrodynamical equations are solved with the help of the Direct
Eulerian Piecewise-Parabolic Method (PPMDE) of Colella \& Woodward
(1984), as implemented in {\sc herakles} solver (Plewa \& M\"uller
2000). Simulations have been done in spherical polar coordinates in a
frame of reference corotating with the protoplanet. {\sc herakles}
guarantees exact conservation of angular momentum which is
particularly important in numerical modeling of disk accretion
problems. The use of its multifluid option with tracer materials
distributed within disk (not presented here) allows to identify the
origin of the material accreted onto protoplanet. The {\sc amra} code
is written purely in FORTRAN 77 and has been successfully used on both
vector supercomputers and superscalar cache-oriented workstations. Its
parallelization on shared memory machines exploits microtasking
(through the use of vendor-specific directives) or the OpenMP
standard.
\section{Simulation setup}
\begin{figure}
\plotfiddle{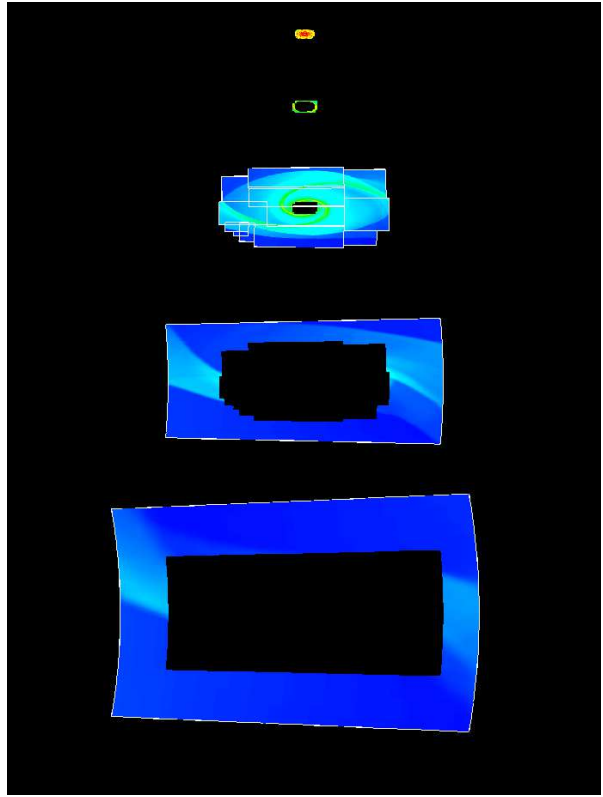}{4in}{0}{40}{40}{-110}{0}
\caption{Breakup of the finest five levels of the refinement hierarchy.}
\label{f:levels}
\end{figure}

\begin{figure}
\plotfiddle{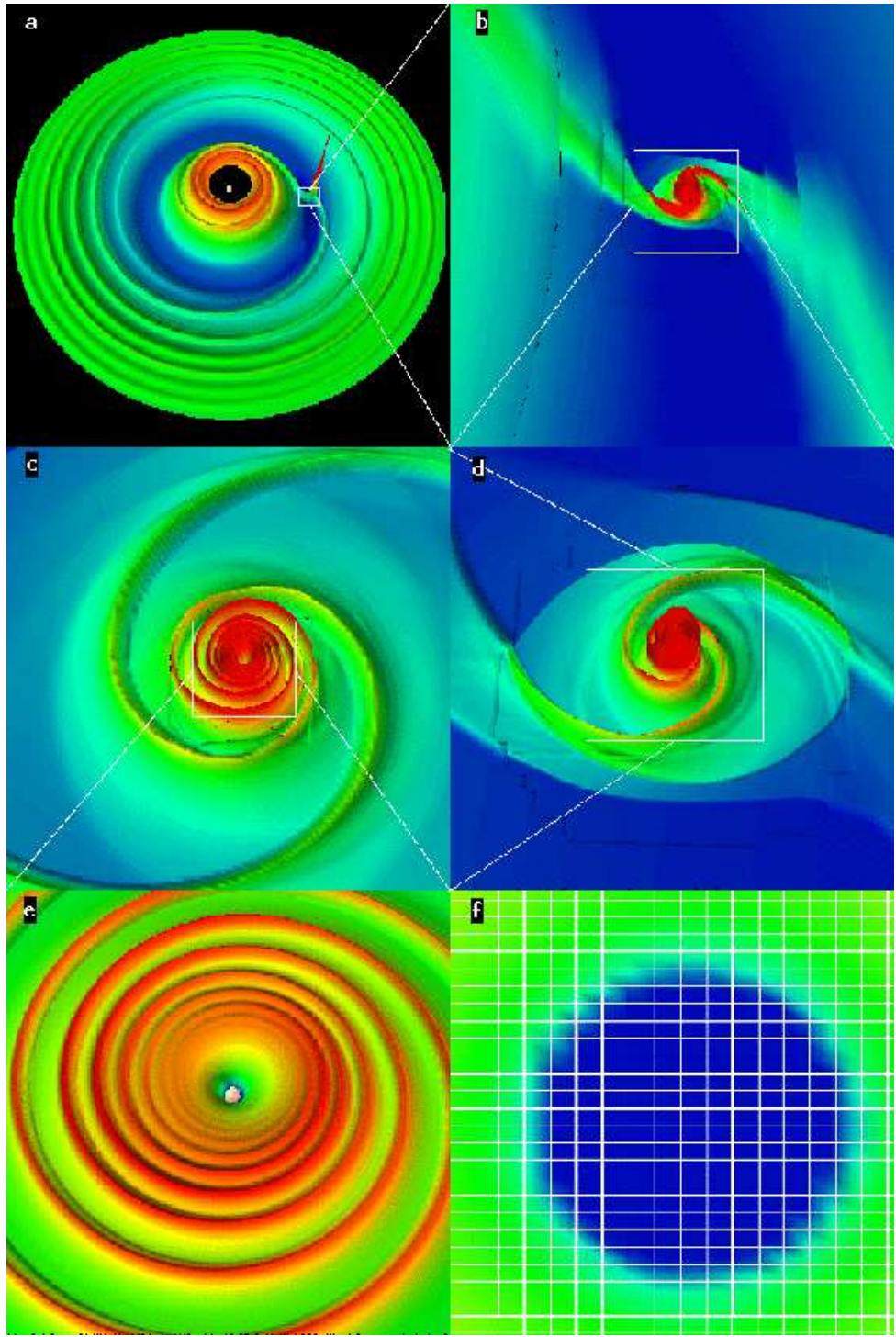}{7.5in}{0}{65}{65}{-170}{0}
\caption{Surface density distribution in the final model (frames (a)-(e)) and the
distribution of grid cells at level 7 in the vicinity of the
planet (frame (f)). There are $\sim 8$ grid cells in the radius of the planet.}
\label{f:zoom-seq}
\end{figure}

The computational domain extends from 0.25 to 2.5 radii of the planet's
orbit.  We employ 7 levels with refinement ratios ranging from (2,4)
to (4,4). The base level contains the protoplanetary (circumstellar)
disk while the 7th level contains the planet and its immediate
vicinity. The base grid consist of $128\times128$ cells uniformly
distributed in $r$ and $\theta$. The effective resolution at the 7th
level is $131072\times524288$ in $r$ and $\theta$, respectively. The
topmost five levels are schematically shown in Figure \ref{f:levels}.
White lines are boundaries of the patches. There are 1, 1, 1, 1, 12, 4
and 49 patches at levels 1-7, respectively. The structure of the grid
at level 7 is shown in Figure \ref{f:zoom-seq}f
with individual cell boundaries drawn with white lines (the dark blue
circle shows size of the planet).
\section{Physical model}
The simulation is initialized with a Keplerian disk. Originally the
disk has a mass of 0.01 M$_{\sun}$, constant $h/r$ ratio of 0.05 and
surface density proportional to $r^{-1/2}$.  The temperature is
a fixed function of $r$ throughout the simulation. There is no explicit
viscosity in the disk. At the outer and inner boundary of the base
grid the gas is allowed to flow freely from the computational
domain. No inflow is allowed for. The accretion onto the planet is
accounted for in a very simplified way.  At every time step the mean
value of the density within two planetary radii is calculated, and
whenever it is higher then a preset value, the excess gas is removed.
At $t=0$ a planet of one Jupiter mass in inserted into the disk on a
circular orbit.  The radius of the orbit and the mass of the planet
remain constant throughout the simulation. The disk is allowed to
evolve for 100 planetary orbits. A gap is cleared in it, and a
secondary, circumplanetary disk is formed.

\begin{figure}
\plotfiddle{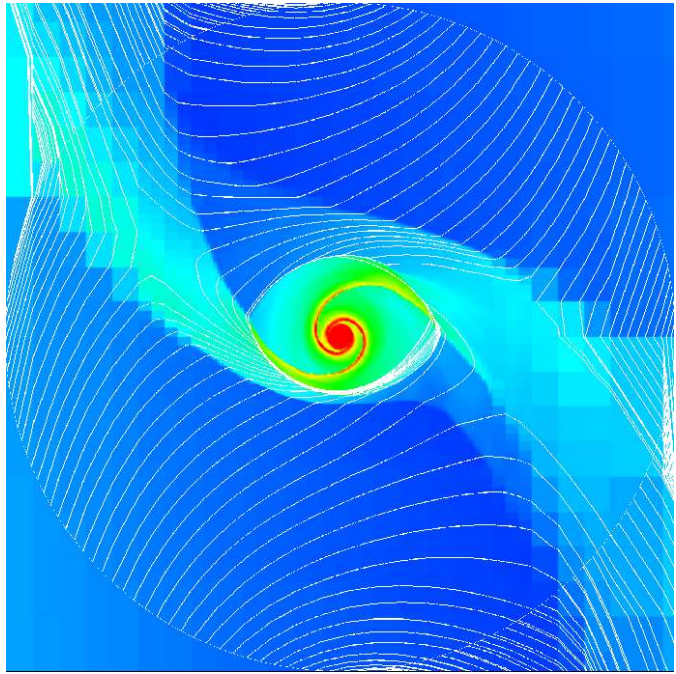}{3.5in}{0}{45}{45}{-130}{0}
\caption{Streamlines of the flow near the circumplanetary disk.}
\label{f:streamlines}
\end{figure}

The sequence of surface plots in Figure \ref{f:zoom-seq} shows the
final structure of both disks (the surface density distribution is
displayed).  The red peak in Figure \ref{f:zoom-seq}a is the
unresolved image of the very dense circumplanetary disk. We have been able
for the first time to see the details of the latter (Figures
\ref{f:zoom-seq}(c)-d).  The streams of gas flowing across the gap from
left and right edge of the frame (light blue) collide with the outer
part of the circumplanetary disk. The collision regions (green wedges) bear
strong resemblance to hot spots in cataclysmic binaries. In every
region two strong shock waves are excited, one of them propagating
into the stream, and the other into the disk. The shocked gas flows
from the collision region along a loosely wound spiral towards the
planet ((Figure \ref{f:zoom-seq}e). This picture is significantly more
detailed than the one recently published by Lubow, Seibert, \&
Artymowicz (2000). Streamlines of the flow around the planet are shown
in Figure \ref{f:streamlines},
and they are in good accordance with those of Lubow et al.

Our simulation is of preliminary nature, and its sole purpose is to
demonstrate the capabilities of {\sc amra}. Currently, we are
improving the physics of the model.  One of the problems we are going
to attack is the calculation of the accurate value of the
gravitational torque from the disk onto the planet in the phase
preceding gap formation.
\section{Visualization}
\begin{figure}
\plotfiddle{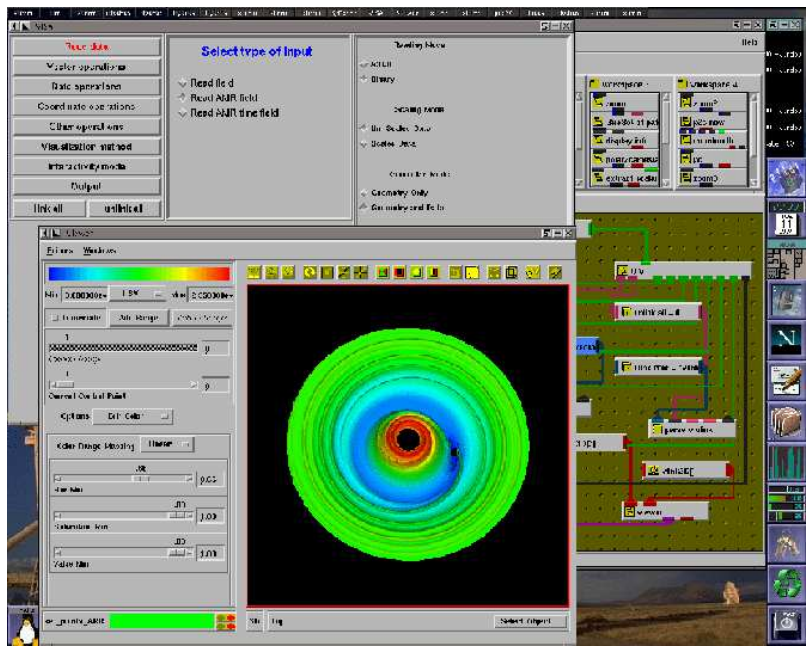}{3.5in}{0}{45}{45}{-160}{0}
\caption{A screenshot of the {\sc visa} application.}
\label{f:visa}
\end{figure}

To visualize the complicated {\sc amra} output, we have chosen the
AVS/Express environment for visual programming. It allows the user to
quickly built simple applications employing standard library
modules. Advanced users can develop their own, highly specialized
modules and applications. Our {\sc amra}-visualization application
({\sc visa}) is partly based on modules written by Favre, Walder, \&
Follini (1999), which have been substantially modified, and partly on
our own modules. A screenshot of {\sc visa} is shown in Figure
\ref{f:visa}.
The panel and the viewer are contained in the two topmost windows,
while the bottom window contains the AVS/Express programming
platform. Currently we are able to read the AMR data, extract
components, perform mathematical operations on data sets and
coordinates, extract any subset of levels or patches, and apply to
them various visualization technique (e.g.\ 2-D plot, surface plot,
isolines, slice). Streamlines can also be calculated. The application
is still under development, and new options are being added.
\acknowledgements{This research is supported by the Polish Committee
for Scientific Research through the grant 2.P03D.004.13.}
\end{document}